# Site-specific Forest-assembly of Single-Wall Carbon Nanotubes on Electron-beam Patterned SiO$_x$/Si Substrates

*Haoyan Wei [a], Sang Nyon Kim [b], Sejong Kim [b], Bryan D. Huey [a], Fotios Papadimitrakopoulos [b], Harris L. Marcus [a]*

[a] Materials Science and Engineering Program, Department of Chemical, Materials and Biomolecular Engineering, Institute of Materials Science, University of Connecticut, Storrs, CT 06269, USA

[b] Nanomaterials Optoelectronics Laboratory, Polymer Program, Institute of Materials Science, Department of Chemistry, University of Connecticut, Storrs, CT 06269, USA

ABSTRACT

Based on electron-beam direct writing on the $SiO_x/Si$ substrates, favorable absorption sites for ferric cations ($Fe^{3+}$ ions) were created on the surface oxide layer. This allowed $Fe^{3+}$-assisted self-assembled arrays of single-wall carbon nanotube (SWNT) probes to be produced. Auger investigation indicated that the incident energetic electrons depleted oxygen, creating more dangling bonds around Si atoms at the surface of the $SiO_x$ layer. This resulted in a distinct difference in the friction forces from unexposed regions as measured by lateral force microscopy (LFM). Atomic force microscopy (AFM) affirmed that the irradiated domains absorbed considerably more $Fe^{3+}$ ions upon immersion into pH 2.2 aqueous $FeCl_3$ solution. This rendered a greater yield of FeO(OH)/FeOCl precipitates, primarily FeO(OH), upon subsequent washing with lightly basic dimethylformamide (DMF) solution. Such selective metal-functionalization established the basis for the subsequent patterned forest-assembly of SWNTs as demonstrated by resonance Raman spectroscopy.

KEYWORDS

Single-wall carbon nanotube forests, metal-assisted self-assembly, electron-beam writing, Si substrates

# 1. Introduction

Carbon nanotubes (CNTs) have attracted considerable attention due to their unique structures, remarkable mechanical and electrical properties, as well as chemical stability [1-4]. They have found applicability to a wide range of electronics applications including nanodevices [5-9], sensors [10, 11] and field emitters [12, 13]. Their unique one-dimensional electronic structures make them ideal molecular wires to transport electrons from and to biological entities, such as peptides [14-18], bound along their length or on their carboxylated ends [19] and therefore have received considerable attention in designing nano-scale biosensors. The unique electron emission properties of vertically oriented nanotubes have attracted the interest in the industry such as flat panel display [13], parallel electron-beam nanolithography and miniature X-ray generators [20]. For a number of these applications, it is very important to be able to produce uniform CNTs with controlled orientation normal to the substrates at specific locations. A number of researchers have reported on growing aligned CNTs using chemical vapor deposition (CVD) on patterned metallic catalyst surfaces achieved via various methods such as standard lithography [21, 22], soft lithography [23], ink-jet printing [24] and nano-channel templates [25, 26]. Although the CVD-grown CNTs possess the right orientation to serve as probes, they are loosely packed (density only around $10^{11}$ cm$^{-2}$) [26], which renders them extremely difficult to be handled in the presence of solvents and upon drying they easily collapse. The co-existence of various types of metallic (*met-*) and semiconducting (*sem-*) single-wall CNTs (SWNTs) could also pose serious problems for electronic devices in which only semiconducting or only metallic nanoprobes are needed [27].

Etching CNTs in an oxidizing environment, such as mixture of sulfuric acid and nitric acid [19] or ozonation [28], presents an alternative viable route to form highly oriented CNT arrays, which could take advantage of post-synthesis SWNTs separation according to length [29] and type (*sem-* versus *met-*) [30]. The shortened carboxylated-nanotubes are anchored to the underlying substrates through either metal-assisted physical adsorption [28] or chemical assembly based on a condensation reaction [31, 32]. Prior research in our laboratory have shown that perpendicular rope-lattice SWNT forests can be readily obtained by assembling nanotubes from dimethylformamide (DMF) dispersion onto an underlying substrate *via* the linkage of $Fe^{3+}$ ions [33]. The resulting density is *ca.* 1.1-1.2 g cm$^{-3}$ ($10^{13}$ cm$^{-2}$) [33], which compares favorably with 1.33 g cm$^{-3}$ density of a van der Waals rope-lattice crystal composed of 1.3-1.4 nm diameter SWNTs [34]. In a recent publication [35], we demonstrated a novel approach for non-thiol functionalization of SWNT forests on gold based on the direct oxidization of Au surface by $Fe^{3+}$ ions as a result of aqua regia effect. This eliminates the time-consuming thiol adsorption process [28, 31, 32], which as well as low surface coverage [36] are among the few drawbacks of conventional thiol functionalization of SWNTs on gold.

Since $Fe^{3+}$ ions act as linkers to bridge SWNTs onto the underlying substrates, the appropriate placement of $Fe^{3+}$ ions is pivotal to achieve patterned SWNT forests. Recently, we demonstrated the patternability of SWNT forest-assemblies on polyelectrolyte films (Nafion) [37] with the aid of electron-beam writing by modifying surface affinity to $Fe^{3+}$ ions. In such patterning process, one factor causing residual iron deposits in the irradiated Nafion films is the enhanced affinity of underlying $SiO_x$/Si substrates to $Fe^{3+}$ ions upon exposure to energetic electrons [37, 38]. On the basis of this finding, we, in the present contribution, investigate the possibility of patterning with low-energy (500 eV) electron-beam directly on Si substrates covered with native oxides ($SiO_x$). As a result of the depletion of oxygen and creation of excess Si dangling bonds, preferential nucleation sites for $Fe^{3+}$ ions are generated within the exposed areas, leading to considerably more SWNT deposits in the subsequent self-assembly.

## 2. Materials and methods

Iron (III) chloride hexahydrate ($FeCl_3 \cdot 6H_2O$, A.C.S. reagent), nitric acid (98%), sulfuric acid (96.4%) and hydrochloric acid (38%) were obtained from Aldrich and used as is. A.C.S. reagent dimethylformamide (DMF) was purchased from J.T.Baker. Hydrogen peroxide (30%) was obtained from Fisher Scientific. Millipore quality deionized water with resistivity larger than 18MΩ was used for all experiments. Si (100) wafers were obtained from Montco Silicon and treated in Piranha solution (concentrated $H_2SO_4$ and 30% $H_2O_2$, 7:3 v/v. *Caution! Piranha solution is a very strong oxidizer and* *reacts violently with organic materials. It should be handled with extreme care*.) at 90°C for 30 min resulting in a clean oxide layer of amorphous $SiO_x$. HiPco SWNTs were purchased from Carbon Nanotechnologies, Inc (CNI). Following the previously established protocol [19, 39, 40], pristine SWNTs were treated in 3:1 mixture of $HNO_3$ and $H_2SO_4$ with sonication for 4 hours at 70°C, filtered, washed with copious deionized water until the pH of filtrated water reached neutral and dried overnight in vacuum. The average length of the resulting SWNTs was *ca.* 25-250 nm [29]. Sonicating these purified shortened-SWNTs (s-SWNTs) in DMF resulted in stable dispersion.

Low-energy (500 eV) electron-beam patterning of $SiO_x$/Si substrates was performed in the PHI Auger 590 system with chamber pressure $10^{-9}$-$10^{-10}$ torr through a 500 mesh TEM copper grid blocking electrons from areas where the subsequent SWNT/$Fe^{3+}$ deposition is not expected (Scheme 1). A Faraday cup was used to measure the electron-beam current to calculate the doses required for exposure.

SWNT forests were assembled using a procedure described previously [33, 41] with the introduction of an extra HCl wash right before exposure to DMF solvent. Briefly, $SiO_x$/Si substrates were first modified with electron-beam using the aforementioned irradiation method. SWNT/$Fe^{3+}$ assemblies were then obtained by sequential dipping of the exposed $SiO_x$/Si substrates into (i) $FeCl_3$ (pH 2.2, 15 min) solution, (ii) a quick wash in aqueous HCl solution (pH < 4) to remove loosely bound $Fe^{3+}$ ions, (iii) a brief wash in DMF (pH 12.7) to remove excess water and facilitate $Fe^{3+}$ ions to transform into their basic hydroxide form [41], and (iv) a 30 min immersion in DMF dispersed SWNTs (pH 8.5) to enable the assembly of SWNT forest arrays.

AFM characterization was conducted on Topometrix Explorer and Asylum Research MFP-3D to investigate the exposed $SiO_x$/Si substrates and the patterns of iron deposits. The Topometrix Explorer employs $Si_3N_4$ AFM probes (Model # MLCT-EXMT-A, spring constant 0.05 N/m, resonant frequency 22 kHz) purchased from Veeco for lateral force microscopy (LFM) imaging in contact mode. The MFP-3D utilizes standard Si AFM probes (Model # AC 160, spring constant 42 N/m, resonant frequency 300 kHz) for topological imaging in AC mode (tapping mode).

The mechanism of electron-beam irradiation on SiO$_x$/Si was studied by Auger spectroscopy investigating the change of oxide layer (SiO$_x$) by monitoring both Si and O Auger lines. The survey was performed in the same Auger system used for the SiO$_x$/Si substrate irradiation (PHI Auger 590 operating at 3 keV with analysis chamber base pressure of *ca.* 10$^{-9}$-10$^{-10}$ torr).

Preferential self-assembly of SWNT forests was characterized with Raman resonance spectroscopy using a Renishaw Ramanscope 2000 equipped with a 514 nm laser source focused on a 1 μm spot by a 100x objective lens.

## 3. Results and discussion

### 3.1. Electron-beam writing on SiO$_x$/Si substrates

After treatment with Piranha solution, a thin layer of chemical oxides (or "native" oxides) SiO$_x$ (x<2, thickness around 0.6 ~ 2 nm) [42] formed on the surface of Si substrates, which is extremely hydrophilic with the measured contact angle of *ca.* 5±1°. This hydrophilicity arises from the silanol groups (Si-OH) terminated on the surfaces, whose concentration is *ca.* 5.0 x 10$^{14}$/cm$^2$ (5.0 /nm$^2$) [43]. This passivation layer has a dangling bond defect density on the order of 10$^{12}$ /cm$^2$ at the SiO$_x$/Si interface, which is about two orders of magnitude higher than that of thermal oxides (stoichiometric SiO$_2$ in the form of siloxane rings Si-O-Si) [42].

Auger spectroscopy is a very valuable tool to characterize the oxidization state of Si substrates due to large shift of the Si peaks. Fig. 1 provides representative Auger spectra of Si LVV (atomic shells, V is valence shell) at different oxide states (thermally grown oxides SiO$_2$, native oxides SiO$_x$/Si and pure Si), which clearly show the complete evolution from pure SiO$_2$ to pure Si. As indicated in Fig. 1a, the lines arising from stoichiometric SiO$_2$ were observed for thermal oxides with major peak centered at 75 eV and two doublets at 58 eV and 61 eV. For Piranha cleaned Si substrates (Fig. 1b), two major peaks (83 eV and 95 eV) were observed for the chemical oxides SiO$_x$/Si. The peak centered at 95 eV originated from the pure underlying Si matrix and the peak centered at 83 eV is believed to come from the SiO$_x$ layer [44], which is the result of the shifting of 75 eV peak in Fig. 1a due to the stoichiometric change from SiO$_2$ to SiO$_x$ (x<2). After sputtering with Ar ions for 30-60 seconds (Ar pressure 14 mPA, emission current 25 mA and emission scale 0.1), the top $SiO_x$ layer was removed and fresh Si surfaces were exposed to the ultra-high vacuum. As demonstrated in Fig. 1c, the peak assigned to $SiO_x$ (83 eV) disappeared and only the Si peak (96 eV) is left.

The Piranha treated $SiO_x$/Si substrates were then transferred to an Auger system for electron-beam patterning through a TEM grid (Scheme 1). Electron-beam of 500 eV was adopted to maximize the beam interaction with top $SiO_x$ layers since low energy electrons (10-500 eV) have small mean free path and will deposit most of their energy onto the surface [45, 46]. After exposure, patterned $SiO_x$/Si substrates were examined with AFM operating in contact mode, measuring both surface topology and friction force. Although no significative topological variation was found (Fig. 2a), clear patterns were observed in lateral force microscopy image as shown in Fig. 2b. Since friction force is representative of surface chemical functionalities [47], the distinct contrast between electron-beam irradiated domains (squares in Fig. 2b) and unirradiated regions (grids in Fig. 2b) indicated that the surface properties of $SiO_x$/Si substrates were transformed upon exposure to energetic electrons.

In order to investigate the possible mechanism of electron-beam irradiation on $SiO_x$/Si, accelerated experiments were performed using Auger spectroscopy to obtain a better insight into the changes of $SiO_x$ layer. Since Auger survey itself (typically conducted at 3 keV) is an irradiation process in point mode with very high dosage (converted area doses were on the order of $10^6$ $\mu C/cm^2$), in contrast the irradiation dose made in rastering mode is so small (typically $10^2 \sim 10^3$ $\mu C/cm^2$) that it could be neglected. Therefore the irradiation and the Auger spectra collection were conducted simultaneously using a stationary electron-beam at 3 keV monitoring the changes of Si LVV peaks and O KVV peaks as a function of electron dosage. In such accelerated experiments, 3 keV was adopted instead of foregoing 500 eV in order to have sufficient signal-to-noise ratio discerning the subtle variation. Fig. 3 showed that the intensity of the peak associated with Si (95 eV) increased while the one assigned to $SiO_x$ (83 eV) decreased. At the same time, oxygen peaks were changing in the same trend as $SiO_x$. This reduction of Auger intensity for $SiO_x$ and O implied that this thin layer of chemical oxide $SiO_x$ was susceptible to the incident electrons. These phenomena originate from the nature that energetic electrons could cause the scission of Si-O bonds, rendering the desorption of oxygen [48]. The change of the irradiated oxide stoichiometry can be schematically represented as following:

$$SiO_x/Si \xrightarrow{e-beam\ irradiation} SiO_y/Si \quad (y < x)$$

The SiOx decomposition onsets were reported to be about 40~120 eV [49]. The Si atoms remaining in the oxide layer were believed to be nonbonding or partially bonding to the matrix [48]. This status indicates that the loss of oxygen atoms creates more dangling bonds around Si atoms in the $SiO_x$ layer. Thus the irradiated regions are expected to be more active than the pristine areas since those Si atoms tend to saturate their dangling bonds by interacting with other species (in our case, absorbing more metal cations *i.e.* $Fe^{3+}$). In addition, a scanning capacitance microscopy (SCM) study [50] demonstrated that these oxygen-deficient Si atoms could also be the defects to act as carrier traps rendering electrical contrast between irradiated and non-irradiated areas, which could also promote the attraction of $Fe^{3+}$ ions.

Although dramatic changes of Auger peak intensity were observed in Fig. 3 for $SiO_x$, which indicated conceivable mass loss might occur, the aforementioned AFM profilometry found no significative topology variation after electron-beam irradiation (Fig. 2a). This fact is attributed to the very different irradiation parameters (irradiation dosage) used for accelerated test and real patterning. In the accelerated experiments, the irradiation was made under stationary electron-beam, whose converted area doses (on the order of $10^6$ μC/cm$^2$) were much higher than that of the irradiation with 500 eV in rastering mode for patterning (typically on the order of $10^2$~$10^3$ μC/cm$^2$). Although oxygen atoms could desorb from the surface as a result of cleavage of Si-O bonds under 500eV electron-beam irradiation, at such low doses only very small fraction of oxygen atoms would lose, which are not expected to be sufficient to cause variations of surface topology due to change in surface oygen concentration..

**3.2. Preferential deposition of FeO(OH)/FeOCl crystallites on electron-beam exposed $SiO_x$/Si domains**

After exposure, the immersion of these electron-beam patterned SiO$_x$/Si substrates into FeCl$_3$ solutions caused differential absorption of Fe$^{3+}$ ions between irradiated regions and unirradiated areas. Excessive FeCl$_3$ was removed from unirradiated areas with an aqueous acidic wash (pH<4) before the samples were exposed onto a basic DMF wash (pH 12.7). Residual surface water content as well as moisture impurities in DMF caused the adsorbed Fe$^{3+}$ ions to precipitate as hydroxides onto these patterned SiO$_x$/Si substrates. Fig. 4a illustrates a typical AFM topological image of iron deposits on electron-beam patterned SiO$_x$/Si substrates. More iron deposits were found in the exposed domains (squares in Fig. 4a). Fig. 4b shows the boundary between the exposed and the unexposed domains. The irradiation domains (on the right) were fully covered with iron precipitates in continuous multi-layers (≥2) configuration. In contrast, only spotty iron precipitates existed in the non-irradiation areas (on the left). Fig. 4c in high magnification indicates that these precipitates appear to have crystalline morphology of size *ca.* 100 and 20 nm in length and diameter respectively [37]. They were identified as FeO(OH)/FeOCl crystallites by TEM characterization with FeO(OH) in majority [37].

Such differential precipitation of iron deposits arose from the distinct affinity to Fe$^{3+}$ ions between non-irradiation regions and electron-beam irradiated areas. As described previously, the wet chemical cleaning (Piranha cleaning) formed a thin layer of chemical oxides SiO$_x$ on the Si substrate surfaces which were terminated with silanol groups (Si-OH). Fe$^{3+}$ cations could chelate to silanol groups through surface complexation reactions. For electron-beam irradiated domains, the affinity to Fe$^{3+}$ ions is greatly

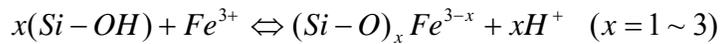

$$x(Si-OH) + Fe^{3+} \Leftrightarrow (Si-O)_x Fe^{3-x} + xH^+ \quad (x = 1 \sim 3)$$

enhanced upon exposure to energetic electrons, which is attributed to the increased dangling bonds of Si atoms as a result of desorption of oxygen atoms. These Si atoms with unsaturated bonds are more active resulting in the enhanced coupling to the Fe$^{3+}$ ions. In addition, the potential of these unbonded Si atoms as carrier traps [50] could also promote the attraction of Fe$^{3+}$ ions through electrostatic interactions.

To minimize the excess Fe$^{3+}$ ions in the non-irradiation regions, an acidic wash was applied to shift the Fe$^{3+}$/Si-OH surface complexation reaction to the left direction by increasing the concentration of product protons (H$^+$) and lowering the concentration of reactant Fe$^{3+}$ ions. Under the same wash conditions, these iron deposits were able to survive on the irradiated regions due to their enhanced affinity to $Fe^{3+}$ ions associated with the damage induced by the electron-beam exposure. The spotty residual iron deposits on unexposed Si domains were believed to mainly originate from the spatial inhomogeneous distribution of silanol groups. If silanol groups were close enough, $Fe^{3+}$ ions could bind more than one silanol group. This multiple bindings hold the $Fe^{3+}$ ions more firmly, and thus make their removal by acidic chemistry very difficult. The irradiation with doses less than 500 μC/cm$^2$ was also attempted and the obtained pattern contrast of iron deposits between the exposed and the unexposed regions was not satisfactory.

**3.3. Site-specific forest-assembly of SWNTs**

The selective metal-functionalization of electron-beam written $SiO_x$/Si substrates provided the basis for the subsequent formation of patterned SWNT forests. Upon immersion of these patterned FeO(OH)-FeOCl/$SiO_x$-Si substrates into DMF dispersed shortened-SWNTs acid-base neutralization between the carboxylic acid terminated nanotubes and the basic iron hydroxides provides the initial driving force for SWNT assembly [33]. The strong hydrophobic interaction between adjacent SWNT side walls was believed to further facilitate the bundling along the lateral direction leading to rope-lattice SWNT forest assemblies. Raman scattering provided the evidence for the site-specific self-assembly of SWNTs on the patterned substrates as shown in Fig. 5 where the characteristic G-band (higher frequency doublet peaks) was only observed in the dark square regions (see the inset optical image in Fig. 5) indicating that the SWNTs were selectively immobilized onto the irradiation areas. The G-band is related to the C-C tangential stretching mode. For SWNTs, the position of the $G^+$ peak (*ca.* 1592 cm$^{-1}$) is dependent on the redox state of SWNTs [51] while the line shape of the $G^-$ peak is largely dependent on nanotube metallicity [52]. In addition, there is also a moderate band observed around 1350 cm$^{-1}$ which comes from the defects in the hexagonal framework of the SWNTs, usually called a disordered-induced band, or D-band [53]. Although some residual iron deposits existed in the unexposed regions (Fig. 4), almost no nanotube Raman signals were detected (Fig. 5). Since carbon nanotube self-assembly was based on electrostatic interactions [33] and such interactions could extend for a few to tens of nanometers [54], the positive polarity of the spotty FeO(OH)/FeOCl crystallites within unexposed regions could be counteracted by the negative polarity of deprotonated silanol groups (Si-O$^-$) surrounding or beneath them. Thus the electrostatic attraction to the negatively charged SWNTs was dramatically reduced, resulting in no or insufficient deposition of SWNT assemblies to give perceivable Raman signals. In contrast, fully covered multilayer FeO(OH)/FeOCl crystallites in the exposed domains assured a clearly positive surface charge that attracted significantly large number of SWNTs.

## 4. Conclusions

The possibility of direct patterning on SiO$_x$/Si substrates using low-energy electron-beam (500 eV) was demonstrated to assist in localizing SWNT forest assemblies. Auger survey indicated that the incident energetic electrons caused the scission of Si-O bonds and desorption of oxygen. Thus, more dangling bonds were created around Si atoms in the oxide layer resulting in an enhanced affinity to Fe$^{3+}$ cations. This led to a distinct difference in the frictional force from unexposed regions as measured by lateral force microscopy. Upon immersion in lightly basic DMF a greater yield of FeO(OH)/FeOCl precipitates in the irradiated areas was affirmed by AFM measurements. The site-specific metal-functionalization established the basis for the subsequent patterned forest-assembly of SWNTs as demonstrated by resonance Raman Spectroscopy. Such spatial localization of nano-structures via electron-beam direct writing could provide a viable route linking novel SWNTs to well-established Si technology in developing biomaterials/nanotube hybrids for biosensor arrays.

**Acknowledgement** The authors gratefully acknowledge the financial support of the U.S. Army Research Office (grant # ARO-DAAD-19-02-1-0381).

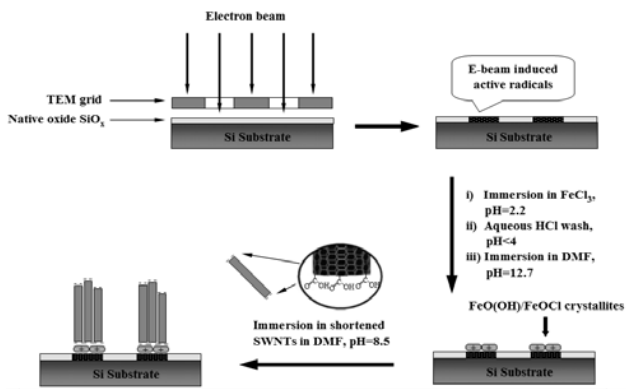

**Scheme 1.** Schematic representation of the formation of SWNT forests on electron-beam patterned $SiO_x$/Si substrates.

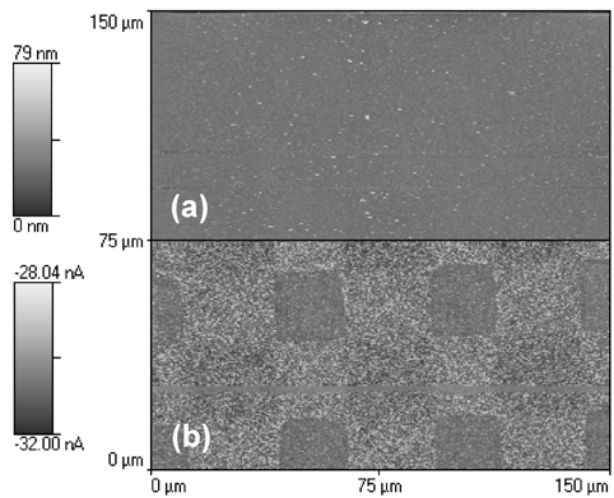

**Fig. 2**. AFM images of island-patterned $SiO_x$/Si substrates irradiated with 500 eV electron-beam and dose of 750 $\mu C/cm^2$. Although no significative variation was found in topological image (a), clear patterns were observed in lateral force microscopy image (b) in which squares were the exposed regions.

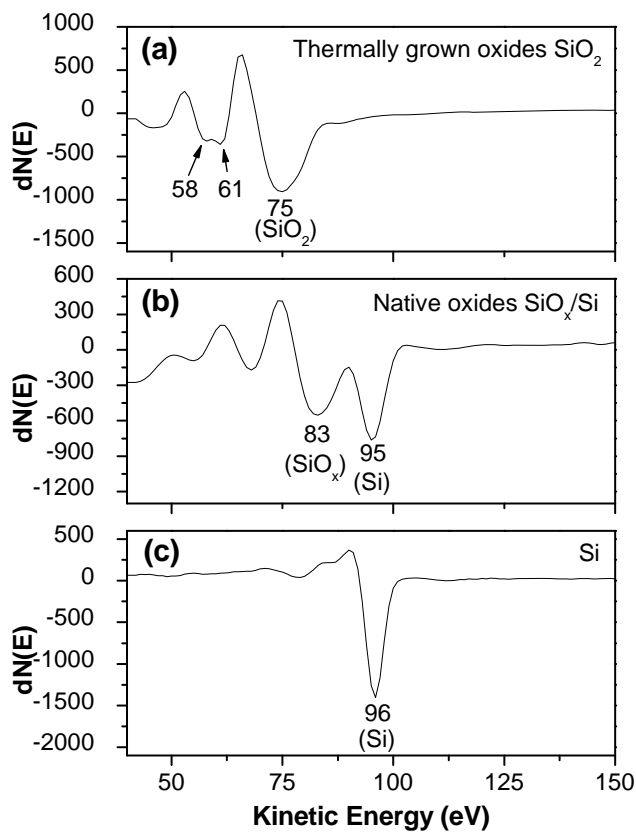

**Fig. 1.** Auger spectra of Si LVV at different oxidization states: (a) thermally grown $SiO_2$, (b) chemical oxide $SiO_x$/Si and (c) Si.

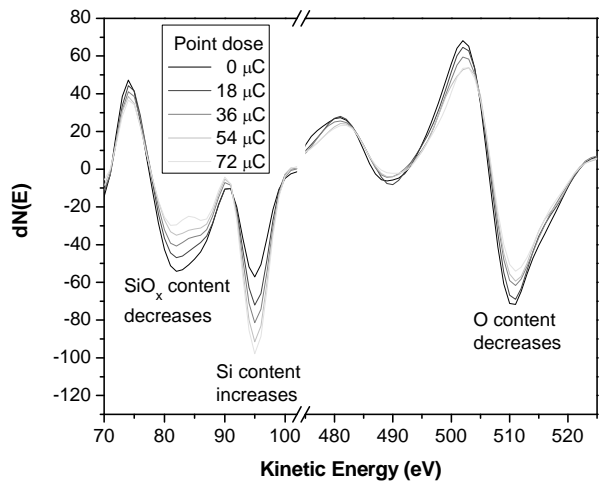
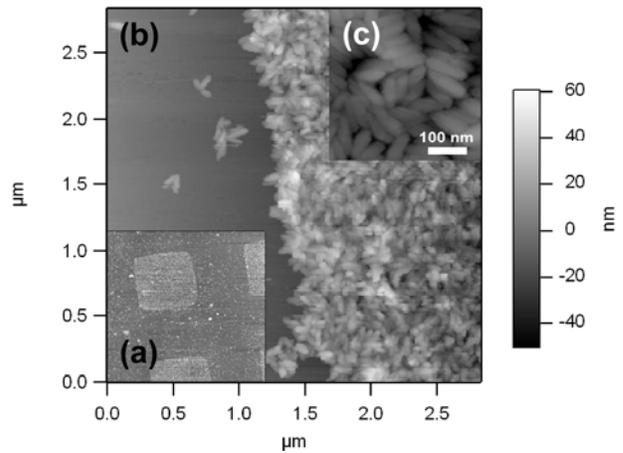

**Fig. 3.** Changes of Si, SiO$_x$ and O Auger lines induced by the electron-beam irradiation. The increase of Si peak intensity and the decrease of SiO$_x$ and O peak intensity implied that oxygen atoms were removed during the irradiation.

**Fig. 4.** (a) Topological AFM image (scan area of 70 x 70 μm$^2$) shows the differential distribution of FeO(OH)/FeOCl crystallites on 500 eV electron-beam patterned SiO$_x$/Si substrates with a dose of 750 μC/cm$^2$. More iron precipitates were found in the exposed regions (squares). (b) The boundary between irradiated (right) and unirradiated regions (left). (c) Morphology of FeO(OH)/FeOCl crystallites in high magnification.

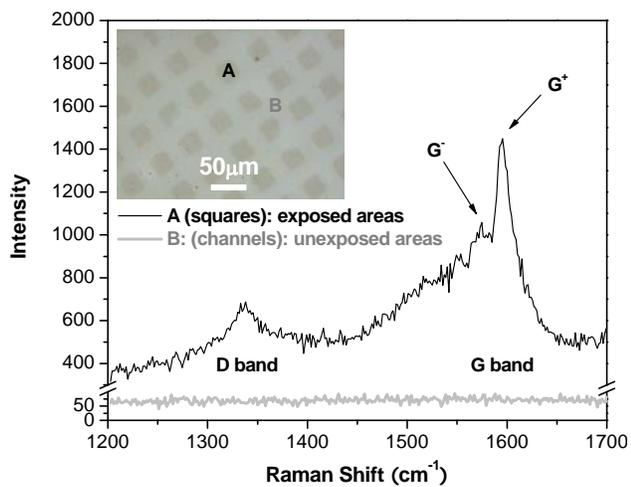

**Fig. 5**. 514 nm Raman spectra of the island-patterned SWNT forests taken at exposed regions (black line) and unexposed regions (gray line) respectively. Inset is the corresponding optical micrograph where dark squares are exposed regions. The $SiO_x$/Si substrate was patterned with 500 eV electron-beam and dose of 1250 $\mu C/cm^2$.